\documentclass[twocolumngrid,prd,superscriptsize,reprint]{revtex4-1}
\usepackage[]{graphicx}
\usepackage{dcolumn}
\usepackage{bm}

\usepackage{amsmath,amssymb,amsfonts}
\usepackage{fancyhdr}
\usepackage{lipsum} 
\usepackage{subcaption}
\usepackage{multirow}

 \usepackage[colorlinks]{hyperref}
 \hypersetup{
     colorlinks = false,
   }

\newcommand{\comment}[1]{}

\usepackage{color}
\usepackage{babel}
\newcommand{\bea}{\begin{eqnarray}}
\newcommand{\eea}{\end{eqnarray}}

\newcommand{\be}{\begin{equation}}
\newcommand{\ee}{\end{equation}}

\begin{document}

\title[]{Constrained instantons and kink-antikink collisions}

\author{C. Adam}
\email[]{adam@fpaxp1.usc.es}
\affiliation{Departamento de F\'isica de Part\'iculas, Universidad de
Santiago de Compostela and \\
Instituto Galego de F\'isica de Altas Enerxias (IGFAE), E-15782
Santiago de Compostela, Spain}

\author{A. García Martín-Caro}
\email[]{alberto.martincaro@usc.es}
\affiliation{Departamento de F\'isica de Part\'iculas, Universidad de
Santiago de Compostela and \\
Instituto Galego de F\'isica de Altas Enerxias (IGFAE), E-15782
Santiago de Compostela, Spain}

\author{M. Huidobro}
\email[]{miguel.huidobro.garcia@usc.es}
\affiliation{Departamento de F\'isica de Part\'iculas, Universidad de
Santiago de Compostela and \\
Instituto Galego de F\'isica de Altas Enerxias (IGFAE), E-15782
Santiago de Compostela, Spain}

\author{K. Oles}
\email[]{katarzyna.slawinska@uj.edu.pl}
\author{T. Romanczukiewicz}
\email[]{tomasz.romanczukiewicz@uj.edu.pl}
\author{A. Wereszczynski}
\email[]{andrzej.wereszczynski@uj.edu.pl}
\affiliation{Institute of Theoretical Physics, Jagiellonian University,
Lojasiewicza 11, Krak\'{o}w, Poland}

\begin{abstract}
We show that constrained $CP^1$ instantons, combined with the Relativistic Moduli Space approach, can accurately describe kink-antikink collisions in the $\phi^4$ model. 
\end{abstract}
\maketitle

\section{Introduction}
Topological solitons are ubiquitous in many areas of theoretical physics reaching from particle theory to nuclear physics, condensed matter or cosmology. A thorough understanding both of their individual properties and their dynamics in multisoliton processes is, therefore, of utmost importance both for classical and quantized solitons. On the other hand, topological solitons typically are solutions of rather complicated, nonlinear field theories, and exact analytical expressions are available only in rare occasions, which considerably complicates their analysis.
One important tool in their study is, therefore, a systematic construction of analytical field configurations which provide an accurate approximation
both for static solitons and for the field configurations which may appear in dynamical processes like soliton scattering.

It turns out that
instantons \cite{Inst-1}, that is, localized, finite action solutions of a field theory in an Euclidean space-time, are particularly useful in this context.
They provide a systematic approach for the understanding of properties of topological solitons in various physically important models. One of the best examples is the Skyrme model, which is an effective theory of quantum chromodynamics in the non-perturbative, low energy regime. This model admits topological excitations, called Skyrmions, which are identified with baryons and atomic nuclei. It was shown by Atiyah and Manton that Skyrmions can be well approximated  by the holonomy of the $SU(2)$ instantons in $(4+0)$ space \cite{AM}. This is important not only because one gets a relatively simple method to generate and study Skyrmions \cite{LM, SSut}, but it is also a way which defines a truncated, finite dimensional configuration space, i.e., a moduli space, in a natural manner, provided by the instanton moduli space (the space of parameters on which a general instanton solution may depend). This finite dimensional subspace is a crucial ingredient both for a simplified study of scattering processes and in the quantization of skyrmions \cite{Ho, Chris-1, chris-2}. Furthermore, the instanton approach allowed for the investigation of dense skyrmionic matter both in the crystal \cite{MS-c, PV-1, M-c} and inhomogeneous phases \cite{PV-2} as well as nucleon-nucleon forces \cite{Chris-3}. It was also used to construct topological solutions of different types like vortices \cite{Nitta} and sphalerons \cite{Sh}. More importantly, a similar construction of skyrmions from instantons becomes realized in some holographic models, such as the Sakai-Sugimoto \cite{SS} or the Sutcliffe \cite{Sut-Inst} models, where Skyrme solitons appear in the boundary theories coupled to an (infinite) tower of vector mesons. 

While applied with success in various areas, the instanton approximation did not work too well in models with massive fields. This is an obvious consequence of the power-like localization of instantons, which is typical for massless fields. 

However, it has recently been shown how to extend this framework also for  Skyrmions with massive pions \cite{Alberto}, where instantons are replaced by the so-called constrained instantons, namely, topologically non-trivial Euclidean configurations whose size has been fixed by an additional constraint in the original action. The effect of such constraint is an exponential decay in the radial direction, which agrees with the solutions of the massive Skyrme model. 

In the present paper, we test the instanton approach for a much simpler model, which is the $\phi^4$ theory in (1+1) dimensions. Therefore we use $CP^1$ instantons. However, our aim is to model one of the most difficult and subtle features of this theory, i.e., the kink-antikink collisions. This goes much beyond the usual application of instantons to model static properties of solitons like, e.g.,  their energies and shapes (geometry).

\section{Instanton approximation }
\subsection{$CP^1$ Instanton and kink profile}

We begin with a short summary of the instanton approximation to kinks in (1+1) dimensions proposed by Sutcliffe \cite{Sut-1, Sut-2, Sut-3}. The obvious choice is to use the $CP^1$ model in (2+0) dimensions, which supports self-dual instantons. They are finite action solutions of the following model 
\be
\mathcal{S}=\int  d^2x \mbox{Tr} (D_\mu Z)^\dagger (D^\mu Z),
\ee
defined on the Euclidean space $x^\mu=(x^1,x^2)$. 
Here $D_\mu = \partial_\mu -A_\mu$, $A_\mu = Z^\dagger \partial_\mu Z$ and $Z$ is a two component complex field obeying a constraint $Z^\dagger Z =1$. The constraint can be resolved via the parametrization 
\be
Z=\frac{1}{\sqrt{1+|w|^2}} \left(
\begin{array}{c}
1 \\
w
\end{array}
\right),
\ee
where $w$ is a complex field. In instanton solutions the complex field $w$ is a rational function of $z=x_1+x_2$. Then, the topological charge of the instanton equals the degree of the rational map $w(z)$. For example, the charge one instanton is given by
\be
w=\lambda(z-z_0)
\ee
where $z_0$ is its position and $\lambda$ its scale, which appear as a consequence of the translational and conformal symmetry. Anti-instantons can be obtained by complex conjugation. 

Now, we define the $U(1)$ abelian holonomy 
\be
U(x_1)= \; \mbox{exp} \left( \int_{-\infty}^\infty A_2 (x_1,x_2) dx_2\right),
\ee
which can be further related to a static scalar field
\be
U(x_1) = e^{i \pi \phi(x_1)}.
\ee
Hence, inserting the single instanton solution we arrive at the following field configuration (from now on, we suppress the lower index, $x\equiv x_1$)
\be
\Phi(x;\lambda, a)= \frac{\lambda(x-a)}{\sqrt{1+\lambda^2(x-a)^2}}.
\ee
So, this profile is independent of the Lagrangian of the scalar field. The only place where the actual kink model may affect the instanton approximation, therefore, is the optimal value of $\lambda$. 

\subsection{$\phi^4$ theory}

In the present work we apply this construction to the $\phi^4$ theory in (1+1) dimensions,
\be
L=\int_{-\infty}^\infty \left( \frac{1}{2} \phi_t^2 - \frac{1}{2}\phi^2_x -\frac{1}{2} (1-\phi^2)^2 \right)dx.
\ee
This model supports a kink $\Phi_K$ and antikink $\Phi_{\bar{K}}$ interpolating between the two vacua $\phi_v=\mp 1$
\be
\Phi^K=\tanh(x-a), \;\;\; \Phi^{\bar{K}}=-\tanh(x-a)
\ee
They are stable, minimal energy solutions in the pertinent topological sectors. Here $a$ is a moduli parameter denoting the center of the soliton. It arises due to the translational invariance of the action and is reflected in the existence of the {\it zero mode}. 

The (anti)kink possesses also a massive mode, so-called {\it shape mode}
\be
\eta^{sh}(x,t)=\sqrt{\frac{3}{2}} \frac{\sinh(x-a)}{\cosh^2(x-a)} e^{i\omega_{sh}t}
\ee
which has a frequency $\omega_{sh}^2=3$. Finally, there are scattering modes, i.e., {\it radiation} for $\omega^2>4$.

To approximate the static $\phi^4$ solitons, one inserts the single instanton into the $\phi^4$ energy integral and minimizes it w.r.t. the scale parameter $\lambda$. The optimal value is $\lambda_0=\frac{2}{\sqrt{3}}$ which leads to the energy $E[\lambda_0]=\frac{\sqrt{3}\pi}{4} \approx 1.36035$. This should be compared with the static energy of the (anti)kink, which is $E_{K(\bar{K})}=\frac{4}{3}\approx 1.33333$. We remark that the observed $2\%$ discrepancy can be made lower if one uses higher dimensional instantons \cite{Sut-2}.

\subsection{Perturbative Relativistic Moduli Space}
Now, we want to go further and model the shape mode of the $\phi^4$ kink. This will be achieved by applying the recently developed perturbative Relativistic Moduli Space approach \cite{AMORW}. The is a version of the collective coordinate method which partially reproduces the Lorentz invariance of the underlying field theory \cite{Rice}, \cite{AMORW}.  

In the collective coordinate model, we assume that for a given solitonic process the infinite dimensional space of all field configurations can be restricted to a finite dimensional set parametrized by some parameters, {\it moduli}, $X^i$
\be
\mathcal{M}[X^i] = \left\{ \Phi(x; X^i); i=1..N \right\}.
\ee 
In the next step, we promote the moduli to time dependent variables. Inserting these configurations into the original Lagrangian and performing the spatial integration brings us to a mechanical like effective model 
\be
L[{\bf X}]=\int_{-\infty}^\infty  \mathcal{L}[\Phi(x; X^i(t))] \, dx
= \frac{1}{2} g_{ij}({\bf X}) \dot{X}^i \dot{X}^j - V({\bf X}) \,,
\label{eff-lag}
\ee
where
\be
g_{ij}({\bf X})=\int_{-\infty}^\infty \frac{\partial \Phi}
{\partial X^i} \frac{\partial \Phi}{\partial X^j} \, dx
\label{modmetric}
\ee
is the metric on $\mathcal{M}$, while
\be
V({\bf X})=\int_{-\infty}^\infty \left( \frac{1}{2}
\left( \frac{\partial \Phi}{\partial x}
\right)^2 + \frac{1}{2}(1-\Phi^2)^2 \right) \, dx
\label{modpot}
\ee
is the effective potential which modifies the motion on the moduli space. 

The Relativistic Moduli Space framework is based on the inclusion of the scale (Derrick) deformation, $x\to bx$. Then, we build the moduli space as follows. We start with energy equivalent set of configurations $\Phi_0(x;a)$. They can be single kink solution, $\phi_K(x-a)$ or some other configurations which are meant to approximate the kink, as, e.g., the instanton motivated profile, $\Phi(x-a; \lambda_0)$. Then, we include a deformation $b$, which rescales the spatial direction, $x \to bx$. Hence, we arrive at the following set of configurations
\be
\mathcal{M}[a,b]= \left\{ \Phi_0(b(x-a)) \right\}
\ee
where we assumed the translational invariance of the problem. We remark that the $\Phi_0(x;a, b=1)$ should be an energy minimizer under the Derrick deformation. 

The resulting CCM possesses a stationary solution $a(t)=vt+a_0, \; b=\gamma \equiv 1/\sqrt{1-v^2}$, which describes a boosted static solution $\Phi_0(x)$,  \cite{Rice}, \cite{AMORW}.

In our case, the Derrick deformation leads to the following family of configurations
\be
\Phi(x; \lambda_0, a,b) = \frac{\lambda_0 b (x-a)}{\sqrt{1+\lambda_0^2 b^2 (x-a)^2}}.
\ee
It is immediately seen that the Derrick deformation can be identified with deformation of the instanton scale parameter $\lambda$. Therefore, configurations with $b=1$ are the energy minima also with the scaling deformation included. 
In addition we can take $\lambda$ as an equivalent scale moduli. Thus, the instanton motivated single kink moduli space reads
\be
\mathcal{M}[a,\lambda] = \left\{ \Phi(x;a,\lambda); \; a\in \mathbb{R}, \; \lambda \in \mathbb{R}_+ \right\}, \label{K-moduli-inst}
\ee 
where the positive sign of $\lambda$ is fixed by the topology, i.e., by the boundary conditions imposed on $\Phi$. 
Obviously, contrary to the flat direction $a$, changes of $\lambda$ correspond to massive deformations which may cover the internal, vibrational modes. As we have already noticed, the corresponding collective model has a stationary solution 
\be
a(t)=vt+a_0, \;\;\; \lambda=\lambda_0 \gamma \equiv \frac{\lambda_0}{\sqrt{1-v^2}},
\ee 
which is just the Lorentz contraction of a boosted instanton approximated kink.

It was shown, however, that to study multi-kink collisions, another version of the Relativistic Moduli Space approach is much more useful. This follows from the observation that for kink-antikink collisions a finite dimensional set of configurations, based on a simple sum of a single kink and antikink, leads to the appearance of an essential singularity on the moduli space at $a=0$, i.e., at the point where solitons are on top of each other and form a vacuum. (Strictly speaking, this only occurs for solitons related by the symmetry $\Phi^{\bar{K}}(x)=-\Phi^K(x)$, which is the case in the $\phi^4$ model.) 

To circumvent this issue, one can use the {\it perturbative} Relativistic Moduli Space approach. This means that we expand the profile provided by the instanton approximation in small perturbations in the scale (Lorentz boost) parameter around the optimal value, $\lambda=\lambda_0+\epsilon$
\bea
\hspace*{-0.8cm} && \Phi_0(x; a, b) =\Phi_0(b(x-a)) = \Phi_0((1+\epsilon)(x-a))  \nonumber \\
&& \quad\quad = \sum_{k=0}^n \frac{\epsilon^k}{k!}(x-a)^k \Phi_0^{(k)}(x-a)
+o(\epsilon^n) \,,
\eea
Now, we treat each term in the expansion as an independent mode, the so-called {\it Derrick mode}
\bea
 \Phi_0(x;a,{\bf C}) &=& \Phi_0(x-a) \nonumber \\
&+& \sum_{k=1}^n \frac{C_{k}}{k!}(x-a)^k\Phi^{(k)}_0(x-a) \,.
\label{K-pert-gen}
\eea
This provides an arbitrarily large set of collective coordinates $C_i, \; i=1..n$, which are the amplitudes of the Derrick modes. 

In the current work we take the simplest possibility and consider only the first, lowest Derrick mode. Hence, we arrive at two dimensional instanton motivated moduli space with configurations of the following form 
\be
\Phi(x;a,C) =  \frac{\lambda_0(x-a)}{\sqrt{1+\lambda_0^2(x-a)^2}} +  C \frac{(x-a)}{(1+\lambda_0^2(x-a)^2)^{3/2}}. \label{pert-K-moduli-inst}
\ee 

\begin{figure}
 \includegraphics[width=0.95\columnwidth]{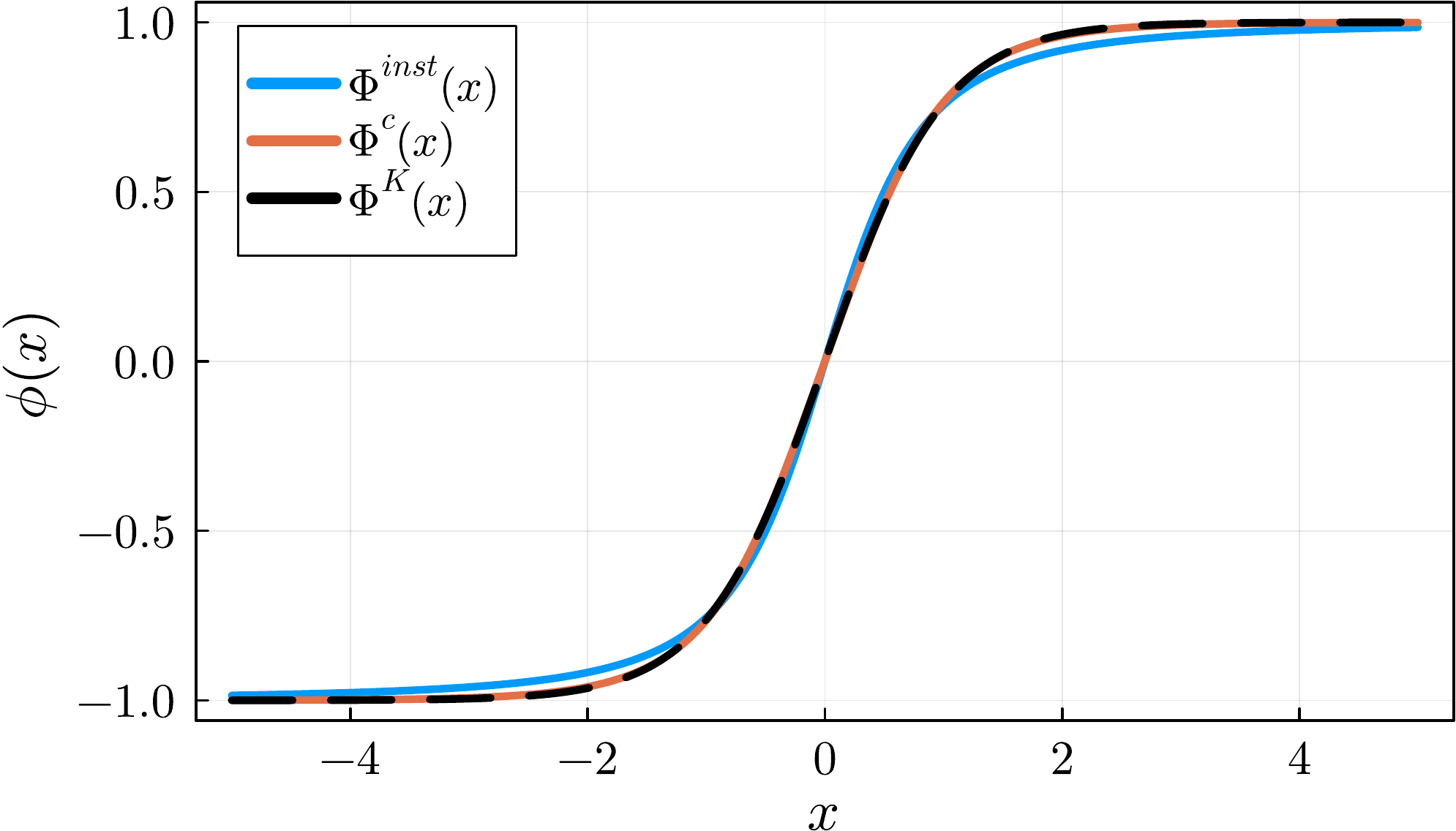}
\caption{Kink profiles: $\phi^4$ kink (black), the instanton approximation (blue) and the constrained instanton approximation (orange).} \label{kink_profiles} 
 \includegraphics[width=0.95\columnwidth]{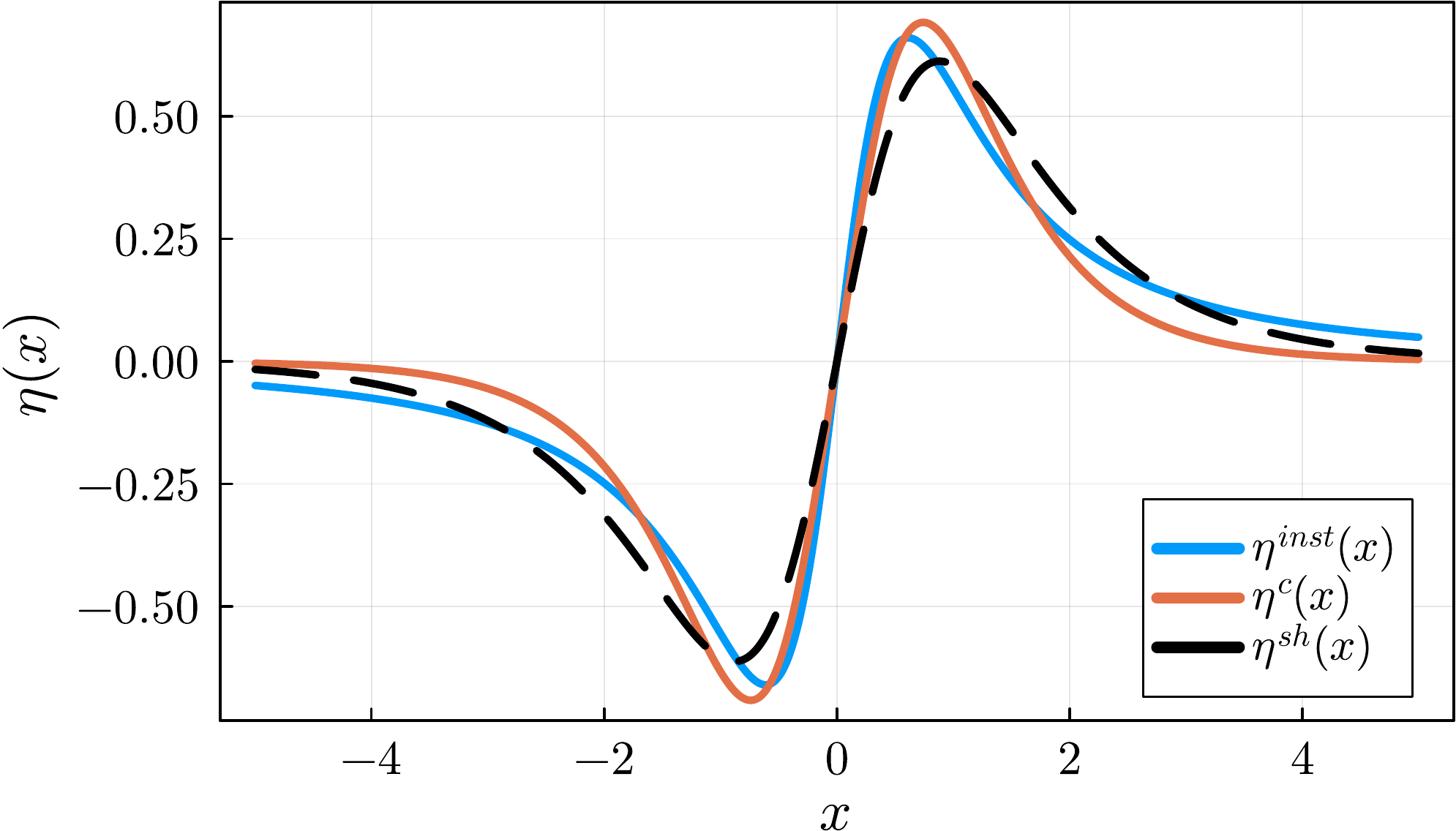}
\caption{Derrick modes compared to the true shape mode of the  $\phi^4$ kink (black): the instanton approximation (blue) and the constrained instanton approximation (orange).} \label{kink_modes}
\end{figure}

It is important to notice that the CCM model arising from these configurations still possesses a stationary solution which approximates the boosted kink. Indeed, inserting $\Phi(x;a,C)$ into the $\phi^4$ Lagrangian we find the following metric
\be
g_{aa}=\frac{\sqrt{3} \pi}{4} + \frac{\pi}{8} C + \frac{9\sqrt{3}\pi}{256} C^2, \;\;\; g_{CC}=\frac{3\sqrt{3}\pi}{64}
\ee
and the effective potential
\be
V(C)=\frac{\sqrt{3} \pi}{4} + \frac{39\sqrt{3} \pi}{512} C^2+ \frac{27\pi}{1024} C^3+ \frac{27\sqrt{3}\pi}{16384}C^4 .
\ee
The resulting equations of motion have a solution 
\be
\dot{a}=v, \;\;\; \tilde{C}=C(v) \label{stat-sol-1}
\ee
where $\tilde{C}$ obeys an algebraic equation 
\be
\frac{v^2}{2} \partial_C g_{aa} = \partial_C V. \label{1-kink-inst}
\ee

Finally, we look closer at the first instanton motivated Derrick mode which after normalization reads
\be
\eta^{inst}= \sqrt{\frac{64}{3\sqrt{3} \pi}} \frac{(x-a)}{(1+\lambda_0^2(x-a)^2)^{3/2}}, \;\;\; \lambda_0=\frac{2}{\sqrt{3}}.
\ee
It is a massive excitation and therefore gives an approximation to the shape mode. It is a rather good approximation. Its frequency is $(\omega_{inst})^2=\frac{13}{4}=3.25$. It has also a very good overlap with the shape mode, $\langle\eta^{inst}|\eta^{sh}\rangle = \int \eta^{sh}\eta^{inst}dx=0.97565.$ 

Hence, we can summarize that the instanton approximation provides a quite good approximation to both static and stationary properties of a kink. Here we mean the mass of the kink and the properties of its normal mode.

One should also be aware that there is an important property of the kink which is not correctly reproduced by the instanton approximations, namely, the near vacuum regime. Indeed, the approximated kinks have very slowly decaying tails. Instead of the original exponential approach to the vacuum
\be
\Phi^{K}=\pm 1 \mp 2 e^{-2x} \;\;\; \mbox{as} \;\;\; x \to \pm \infty,
\ee
the instanton approximation leads to $x^{-1}$ localized solitons. Hence, the approximated kinks interact much more stronger and at a much larger distance. As we will see below, this rather strongly affects the multi-kink dynamics. 
\section{Instanton-antiinstanton valley }
In order to study multi-soliton solutions, one has to find multi-instanton configurations in the corresponding topological (instanton) charge sector. Such $CP^1$ multi-instantons are exactly known. They are given by holomorphic rational maps whose degree is the instanton charge.  Then, for example, a kink-kink scattering may be approximated by a path in the two-instanton space, where the actual time evolution would arise from the pertinent collective coordinate model \cite{Sut-1, Sut-3}. 

There is, however, a problem if we want to study kink-antikink collisions. There are no $CP^1$ instantons in trivial charge sector. Technically it means that there are no nontrivial self-dual solutions with zero topological charge. They may appear if we couple the $CP^1$ model with a background field, an {\it impurity}, in a very specific manner, the so-called BPS impurity models \cite{BPS-imp-1, BPS-imp-2}. Here we will not follow this strategy but, rather, will use a more traditional instanton-antiinstanton ($I\bar{I}$) valley construction \cite{BY}. 

In this construction, one focuses on trajectories $\Psi_v(x^i; \xi)$ in functional space which asymptotically, for a parameter $\xi \to \infty$, represent a pair of infinitely separated instanton and antiinstanton solutions of the action $S_0$. For finite $\xi$, the configurations $\Psi_v(x^i; \xi)$ are not extrema of the original action. Instead, $\Psi_v(x^i; \xi)$ are solutions of a reduced theory which is the original $S_0$ under a constraint that only variations orthogonal to the direction $\xi$ are taken into account
\be
\frac{\delta S_{red}}{\delta \Psi} [\Psi_v] = \omega_\Psi (x^i; \xi) \frac{\partial \Psi_v}{\partial \xi}
\ee
equipped with above mentioned boundary condition. In other words, we find a streamline in the functional space and the valley configuration goes along this streamline.

Examples of such $I\bar{I}$ valleys are know \cite{BW, Dorey, Elis}.  For instance, a valley can have the following simple form \cite{Elis}
\be
w(z,\bar{z}) = \frac{(\xi -1/\xi)\bar{z}}{1+|z|^2}
\ee
This is a concentric $I\bar{I}$ pair which can be transformed into a non-concentric one by a conformal transformation. 

In the present work, we take the simplest choice relevant for kink-antikink collisions and assume the valley in a form which leads to the simple kink-antikink superposition \cite{BW}. We do not know an exact form of the corresponding $I\bar{I}$ configurations but this is not needed for our computations. 
 
After this reasoning we propose the instanton motivated kink-antikink moduli space to be a sum of single soliton cases. Specifically, the relevant configurations are
\bea
&& \Phi^{K\bar{K}} (x; a, C) =   \\
&=& \frac{\lambda_0(x+a)}{\sqrt{1+\lambda_0^2(x+a)^2}}-\frac{\lambda_0(x-a)}{\sqrt{1+\lambda_0^2(x-a)^2}}-1\nonumber \\
&+& C \left( \frac{x+a}{(1+\lambda_0^2(x+a)^2)^{3/2}} -\frac{x-a}{(1+\lambda_0^2(x-a)^2)^{3/2}}\right), \nonumber
\eea
where we keep only the first Derrick mode (symmetrically excited). However, there is still a problem, since the moduli space metric of the resulting CCM (\ref{eff-lag}) has a singularity at $a=0$. Indeed, $g_{CC}=g_{aC}=0$ at this point. Fortunately, this is an apparent singularity which can be removed by a suitable choice of the moduli space coordinates. Here, it is enough to change $C\to \frac{C}{\tanh(a)}$, 
see \cite{MORW-moduli} and \cite{AMORW} for details. This finally brings us to the following set of configurations 
\bea
&& \Phi^{K\bar{K}} (x; a, C) =  \label{KAK-inst} \\
&=& \frac{\lambda_0(x+a)}{\sqrt{1+\lambda_0^2(x+a)^2}}-\frac{\lambda_0(x-a)}{\sqrt{1+\lambda_0^2(x-a)^2}}-1\nonumber \\
&+& \frac{C}{\tanh(a)} \left( \frac{x+a}{(1+\lambda_0^2(x+a)^2)^{3/2}} -\frac{x-a}{(1+\lambda_0^2(x-a)^2)^{3/2}}\right). \nonumber
\eea

The final step is to identify the appropriate initial conditions which would correspond to kink-antikink scattering in the original $\phi^4$ theory. Obviously, for a large initial separation, $a(t=0)=a_0 \gg 1$ the configurations represent free solitons. Thus, the initial conditions can be read of from the single kink sector. Specifically, they read
\be
a(0)=a_0, \;\;\; \dot{a}(0)=v, \;\;\; C(0)=\tilde{C}, \;\;\; \dot{C}(0)=0. \label{CCM-init}
\ee
where $\tilde{C}$ obeys equation (\ref{1-kink-inst}). 
\begin{figure}
 \includegraphics[width=0.95\columnwidth]{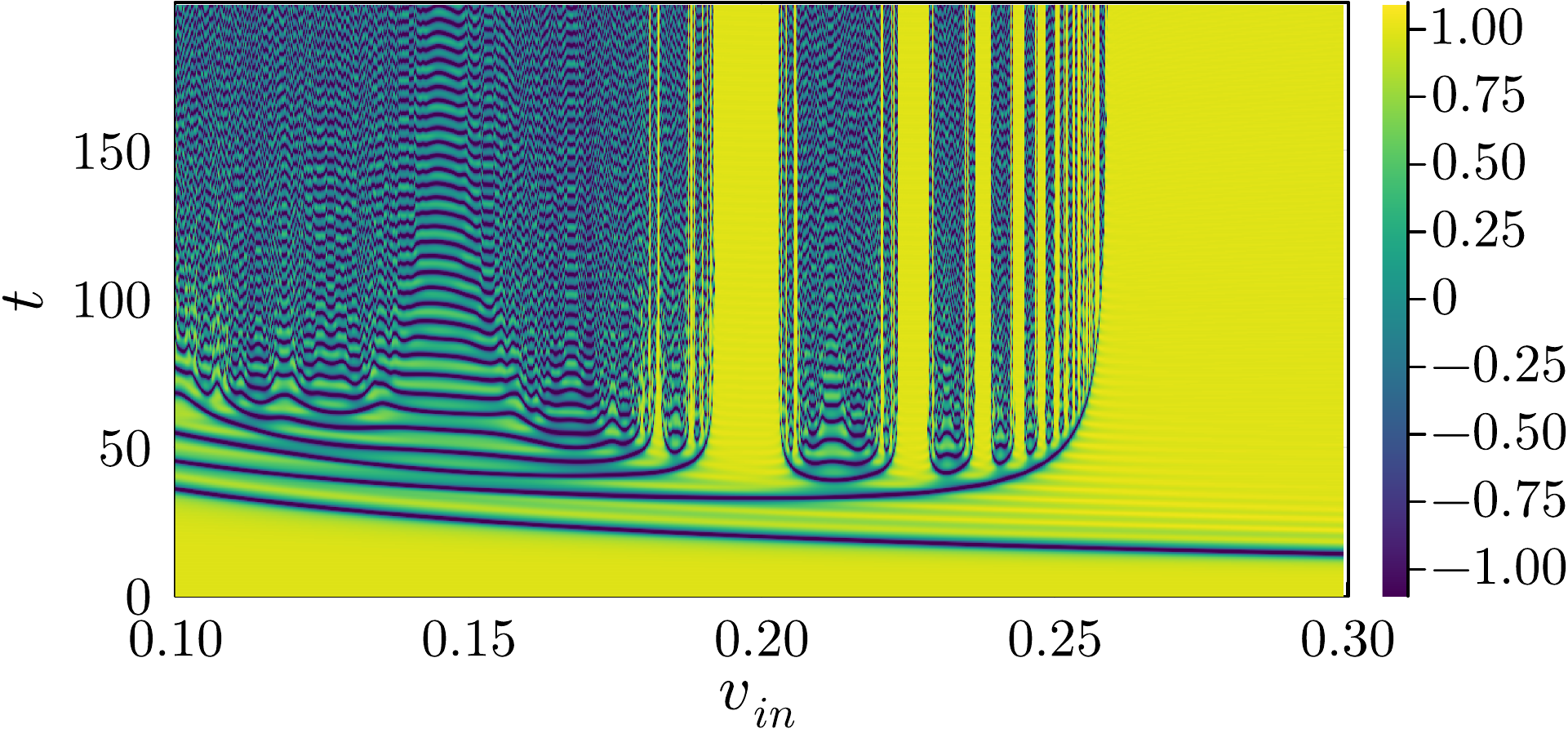}
 \caption{Kink-antikink collision in the $\phi^4$ model.} \label{KAK-phi4-plot}
 \includegraphics[width=0.95\columnwidth]{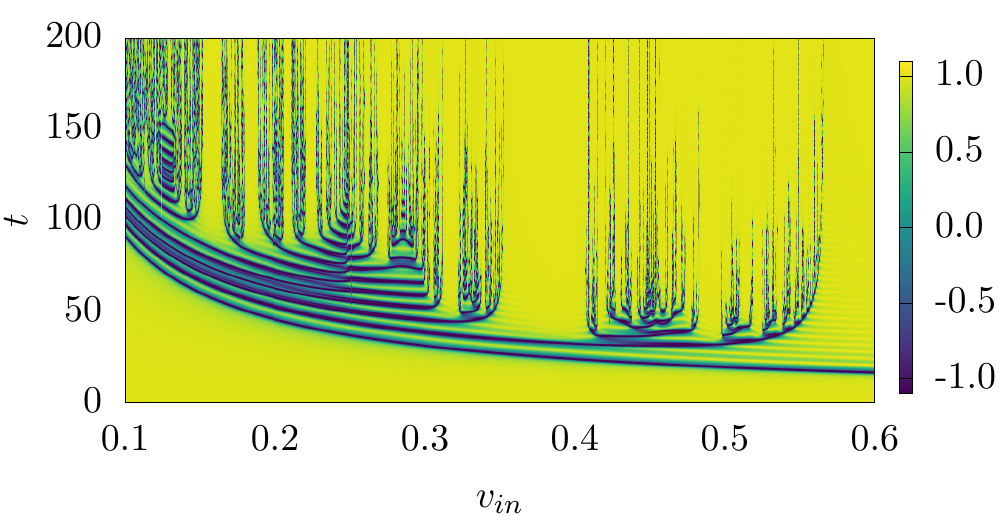}
 \caption{Kink-antikink collision in a CCM based on the superposition of instanton and antiinstanton approximated configurations (\ref{KAK-inst}).}\label{instanton-sum}
   \includegraphics[width=0.95\columnwidth]{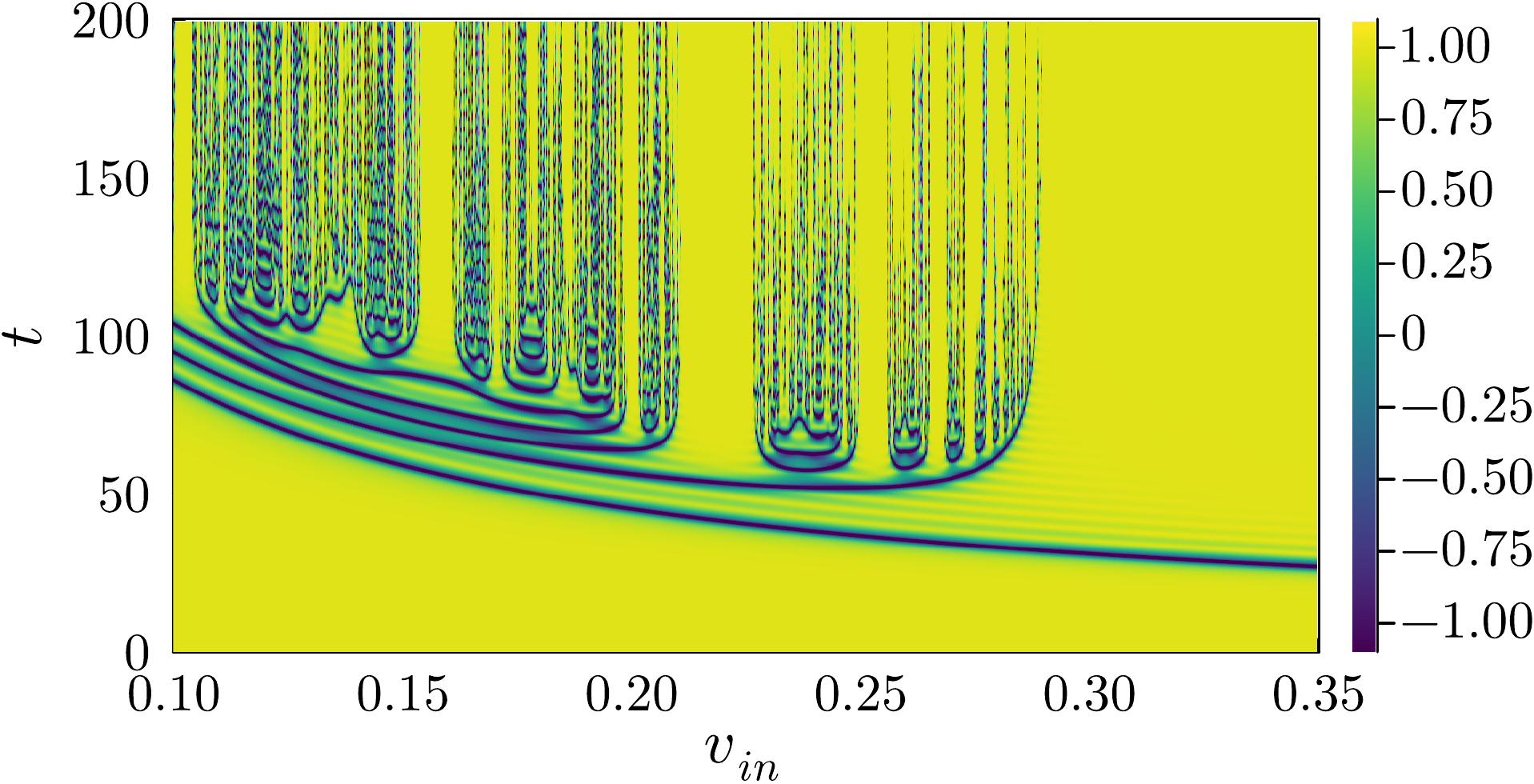}
 \caption{Kink-antikink collision in a CCM based on the superposition
 of constrained instanton and antiinstanton approximated configurations (\ref{kak-c-inst}.} \label{instanton-constr}
 \end{figure}

We start with presenting the results of kink-antikink scattering in the $\phi^4$ model, see Fig. \ref{KAK-phi4-plot}. Here we show the time dependence of the field at the origin $\phi(x=0,t)$ for different initial velocities of the colliding solitons $v \in [0.1,0.3]$. One can easily recognize the bounce windows, where kink and antikink scatter back to infinity after a certain number of collisions, as well as the bion chimneys where the solitons do not have sufficient kinetic energy to overcome the attraction and, therefore, annihilate via the formation of a bion i.e., an oscillating state which slowly decays to the vacuum via radiation. The actual behaviour strongly depends on the initial velocity and reveals a fractal like pattern. For $v>v_{crit}\approx 0.2598$ we see only one-bounce scattering. 

In Fig. \ref{instanton-sum} we show the value of the field at the origin found in the instanton motivated CCM based on configurations (\ref{KAK-inst}). We see bounce windows and bion chimneys, but the critical velocity is very large, more than twice as big as in the true collisions, $v_{crit}^{inst} \approx 0.57$. This means that, although the static and stationary properties of single kink are quite well reproduced by the instanton approximation, it is too little to cover dynamical features of the model as e.g., the effects of kink-antikink collisions. 

An obvious source of this failure is the wrong asymptotic behaviour of the approximated kinks. They decay too slowly and feel each other much stronger than in the original field theory. Therefore, it is not too surprising that the kinks need significantly bigger energy to overcome this stronger attraction. This energy can only come from the initial kinetic energy. As a consequence, we find a much larger critical velocity. 

To improve the instanton description of kink-antikink collisions we have to modify the behaviour of the solitons in the near-vacuum regime and equip the kink and antikink with exponential tails. This can be done by making use of the constrained instantons. 

\section{Constrained instanton}
In order to impose an exponential decay on the (constrained) instanton configurations, we may modify the original instanton Lagrangian by including a mass term in the action
\begin{equation}
    S_m=\frac{1}{2}\int d^2x [\partial_\mu n^a\partial_\mu n^a+m^2(1-n^3)].
\end{equation}
Here we use the $O(3)$ formulation of the model, where instead of the complex field $w$ we have a unit three component iso-vector $\vec{n}=(n^1,n^2,n^3)$
\begin{equation}
    w=\frac{n_1+i n_2}{1-n_3} .
\end{equation}
It will be also useful to introduce polar coordinates in the base manifold, $\{r,\theta\}$, so that an axially symmetric configuration can be easily written in terms of a radial function $f(r)$ as
\begin{equation}
    w(r,\theta)=\frac{\sin f e^{i\theta}}{1-\cos f}.
\end{equation}
For instance, the single-instanton solution centered at the origin is given by ${\cos f(r)=(\lambda^2-r^2)/(\lambda^2+r^2)}$. 

The massive $O(3)$ model does not support topologically nontrivial euclidean action minimizers. However, it can still provide very useful so-called constrained instanton configurations \cite{const-inst1, const-inst2}.

The equation of motion in the large $r$ limit simplifies to the modified Bessel equation,
\begin{equation}
    f''+\frac{1}{r}f'-\left(\frac{1}{r^2}+m^2\right)f=0
\end{equation}
so that solutions vanishing at infinity must satisfy
\begin{equation}
    f\sim K_1(mr)\qquad \text{for}\quad r\rightarrow \infty.
\end{equation} 
On the other hand, we want our constrained instanton configuration to behave as the pure $CP^1$ instanton for small distances from its center,
\begin{equation}
    f\sim \arccos{\frac{\lambda^2-r^2}{\lambda^2+r^2}}\qquad \text{for}\quad r\sim0.
\end{equation}
A simple function that achieves both conditions is
\begin{equation}
    f_c(r)=\arccos\left({\frac{\lambda^2-r^2}{\lambda^2+r^2}}\right)mrK_1(mr)
\end{equation}
This function depends on two parameters $\lambda$ and $m$, which for the moment we leave as free parameters. 

Now, we compute the holonomy of $A_2$ generated by the constrained instanton profile and find
\begin{widetext}
\begin{equation}
    \Phi^c(x; \lambda, m)=\frac{1}{\pi}\int dt\frac{x}{x^2+t^2}\cos ^2\left(\frac{m}{2}  \sqrt{x^2+t^2} K_1\left( m\sqrt{x^2+t^2}\right)\arccos\left(\frac{2\left(x^2+t^2\right)}{\lambda ^2+x^2+t^2}-1\right)\right)
\end{equation}
\end{widetext}

Then, we insert the approximated kink profile into the energy integral of the $\phi^4$ model and find the optimal values of the parameters. They are  $\lambda_0=1.4076$ and $m_0=0.5870$. This guarantees that the scale perturbation will not provide a lower energy configuration. Note that for the constrained instanton we have to perform such a two parameter minimization because for such an ansatz the scaling deformation is not simply equivalent to a redefinition of  $\lambda$. Instead, it affects terms where $\lambda$ is absent. This is due to the explicit conformal symmetry breaking implied by the inclusion of the mass term. We also remark that these values of the ansatz parameters give a very good approximation of the true kink energy. Namely  $E=1.3337$. One can also observe that the optimal value of the mass parameter $m_0$ is not too different from the exponentially decaying tail of the kink, for which $m=1$. 

Of course, the reason why we have to consider scale deformations is clear from the previous analysis. We need Derrick modes to perturbatively approximate the Lorentz boost of a static kink configuration and, therefore, to correctly define the initial conditions in the CCM. 

Thus, the simplest perturbative relativistic moduli space describing single-kink sector is built out of the following configurations
\be
\Phi^c(x-a; \lambda_0,m_0) + C (x-a) \partial_x \Phi^c(x-a;\lambda_0,m_0),
\ee
where we contain only the first Derrick mode (for simplicity, we show it without normalizing to one)
\be
\eta^c= (x-a) \partial_x \Phi^c(x-a;\lambda_0,m_0) .
\ee
Here, $a$ is the position of the kink and $C$ is the amplitude of the first Derick mode. The constant velocity motion of the kink is found as a stationary solution of the resulting CCM, $\dot{a}=v$, $C=\tilde{C}(v)$ obeying eq. (\ref{1-kink-inst}). 

The construction of the kink-antikink moduli space is also straightforward. It is based on the superposition of the kink and antikink configuration with the previously discussed regularization. Hence, 
\bea
\Phi_{K\bar{K}}^c &=&\Phi^c(x+a; \lambda_0,m_0) - \Phi^c(x-a; \lambda_0,m_0) \nonumber \\
&+& \frac{C}{\tanh(a)} \left((x+a) \partial_x \Phi^c(x+a;\lambda_0,m_0) \right. \nonumber \\ 
&-& \left. (x-a) \partial_x \Phi^c(x-a;\lambda_0,m_0) \right) \label{kak-c-inst}
\eea
This gives rise to a CCM which we solve with the previously specified initial conditions (\ref{CCM-init}). 

Solving the equations arising from this CCM, 
\begin{equation}
    \ddot{X}^i=-\Gamma^i_{\,jk}\dot{X}^j\dot{X}^k-g^{ij}\frac{\partial V}{\partial X^j},
\end{equation}
requires calculating many integrals numerically, especially because even the instanton profile is given by a slowly converging integral. Here, the Christoffel symbols and the gradient are functions of the position on the moduli space and are given by two dimensional integrals (over $t$ and $x$). In our earlier attempts, we calculated all the integrals (\ref{modmetric}), (\ref{modpot}) at each time step. However, this approach resulted in many artefacts and substantial, but acceptable computation times. Here, calculating integrals at each time step would be much more time-consuming. Therefore we applied a different strategy. Prior to solving the effective equation, all elements of $\Gamma^i_{\,jk}(X)$ and $\partial^iV(X)$ were calculated on a grid and two dimensional splines were used to approximate the rhs of the effective equations. This reduced the computation time of the full scan shown in Fig. \ref{instanton-constr} down to a couple of minutes (in Julia programming language).

In Fig. \ref{instanton-constr}, we plot the results of kink-antikink collisions in the CCM derived above. Again we see bounce windows and bion chimneys, but now in very good agreement with the full field theory. Especially the critical velocity improves significantly. Now it reads $v_{crit}^{c}= 0.2864$, which is only $10\%$ bigger than the true value. Note that in the usual instanton approximation the difference is about $120\%$. The improvement is spectacular and indicates a high importance of the correct localization, that is, the correct approach to the vacuum of the approximated soliton.

We remark that the results of the constrained instanton based CCM are quite similar to computations where the exact kink and the shape \cite{MORW} or Derrick mode \cite{AMORW} were used. One can verify that the corresponding scans of the collisions are qualitatively similar, while the critical velocities read $v_{crit}=0.282$ and $v_{crit}=0.2853$, respectively.

\section{Conclusions}

In the present paper we show that the constrained instanton approximation  combined with the instanton-antiinstanton valley construction and the perturbative relativistic moduli space approach can quite accurately describe such a delicate phenomenon as kink-antikink collisions in the $\phi^4$ model. This is a very nontrivial result, as the kink-antikink collisions lead to a chaotic system with a fractal structure in the final state formation. 

Our results provide clear evidence that the (constrained) instanton approximation gives a universal tool allowing for the analysis (and/or approximation) of both individual soliton properties and soliton collisions.
Of course, we cannot expect that this approximation will be competitive 
with more specific, taylor-made approaches whenever such a specific treatment can be found like, e.g., the moduli space approach based on a superposition of true single soliton solutions combined with single kink normal modes \cite{MORW}, single kink Derrick modes \cite{AMORW} or even multi-kink internal modes \cite{phi-6}. This is because the instantons provide a limited set of configurations which is basically independent of the theory we want to approximate. The only place where this theory affects the instanton approximation is via the optimal values of the parameters. However, in many cases as for example the Skyrme model, a good, i.e., physically motivated choice for a moduli space and the related CCM is difficult to find, and precisely in these situations the instanton approximation can be a very useful tool.

Therefore, the main result is the demonstration that the constrained instanton approximation can be applied to much more complicated phenomena or properties than the usually considered shapes and masses of static solitons. Specifically, dynamical properties which involve the internal modes seem to be quite well captured. 

This can also be very important for the application of instantons to the analysis of the semiclassical version of solitonic models, where normal modes are the crucial ingredients \cite{Q1}. See also  \cite{Q2, Q3, Q4} for recent developments. Especially, in higher dimensions the instanton approach can provide a simple but trustworthy approach.

\section*{Acknowledgements}

The 
authors acknowledge financial support from the Ministry of Education, Culture, and Sports, Spain (Grant No. PID2020-119632GB-I00), the Xunta de Galicia (Grant No. INCITE09.296.035PR and Centro singular de investigación de Galicia accreditation 2019-2022), the Spanish Consolider-Ingenio 2010 Programme CPAN (CSD2007-00042), and the European Union ERDF.
 AGMC is grateful to the Spanish Ministry of Science, Innovation and Universities, and the European Social Fund for the funding of his predoctoral research activity (\emph{Ayuda para contratos predoctorales para la formaci\'on de doctores} 2019). MHG thanks the Xunta de Galicia (Consellería de Cultura, Educación y Universidad) for the funding of his predoctoral activity through \emph{Programa de ayudas a la etapa predoctoral} 2021. KO was supported by the Polish National Science Centre 
(Grant NCN 2021/43/D/ST2/01122). 

AW thanks Christopher Halcrow for comments.

\end{document}